\documentclass[superscriptaddress,amsmath, nofootinbib]{revtex4}
\usepackage{amsfonts}
\usepackage{graphicx}
\usepackage[latin1]{inputenc}
\usepackage{amsmath}
\usepackage{amssymb}
\usepackage{euscript}
\usepackage{hyperref}
\begin{document}


\title{Variational problem and Hamiltonian formulation of the Lagrange-d'Alembert equations with nonlinear nonholonomic constraints.}

\author{Alexei A. Deriglazov }
\email{alexei.deriglazov@ufjf.br} \affiliation{Depto. de Matem\'atica, ICE, Universidade Federal de Juiz de Fora,
MG, Brazil} 

\author{}

\date{\today}

\begin{abstract}
Any given system of ordinary differential equations in $n$\,-dimensional configuration space can be obtained from a peculiar variational problem with one local symmetry. The obtained action functional leads to the Hamiltonian formulation in $4n+2$\,-dimensional phase space. As concrete examples, we discuss the cases of Lagrange-d'Alembert equations with nonlinear nonholonomic constraints, as well as the equations of motion with dissipative (frictional) forces.
\end{abstract}

\maketitle 



\section{Introduction. Lagrange-d'Alembert equations.}

Nonholonomic constraints arise not only in classical problems of rigid body dynamics \cite{Routh_1897, Cha_2002, Cha_2008}, but also in current applications, where velocity-dependent constraints are typical in the description of rotating bodies and spinning particles in both relativistic and generally-covariant theories \cite{Ili_2025, Dju_2025, Ote_2025, Kim_2025, Kho_2024, Spi_2017, Kun_2025, Cao_2024, Har_2025, Ray_2025}.  The task of incorporating such constraints into the general framework of Lagrangian and Hamiltonian formalism is an actual issue, which is discussed in the present note based on the Lagrange-d'Alembert equations.

To avoid unimportant technical details, we consider the simple Lagrangian action 
\begin{eqnarray}\label{nh1}
S_0=\int dt ~ \frac12\dot {\boldsymbol q}^2-U({\boldsymbol q}),  \quad \mbox{this implies} \quad \ddot q^i+\frac{\partial U({\boldsymbol q})}{\partial q^i}=0,
\end{eqnarray}
in the configuration space with generalized coordinates ${\boldsymbol q}=(q^1, q^2, \ldots, q^n)$. We assume that the coordinates are subject to a single nonholonomic, nonlinear, and nonintegrable constraint $D$
\begin{eqnarray}\label{nh2}
D({\boldsymbol q}, \dot {\boldsymbol q})=0, \qquad \sum_i \left(\frac{\partial D}{\partial \dot q^i}\right)^2>0, \qquad 
D({\boldsymbol q}, \dot {\boldsymbol q})\ne\frac{dG({\boldsymbol q})}{dt} \quad \mbox{for some} ~  G({\boldsymbol q}).  
\end{eqnarray}
It is known that a self-consistent evolution of such a system can be obtained from the following Cauchy problem for the Lagrange-d'Alembert equations:  
\begin{eqnarray}
\ddot q^i+\frac{\partial U({\boldsymbol q})}{\partial q^i}-h\frac{\partial D({\boldsymbol q}, \dot {\boldsymbol q})}{\partial \dot q^i}=0, \label{nh3} \qquad \qquad \qquad \qquad \\
{\boldsymbol q}(0)={\boldsymbol q}_0, \qquad \dot{\boldsymbol q}(0)=\dot{\boldsymbol q}_0, \qquad 
\mbox{such, that} \quad D({\boldsymbol q}_0, \dot {\boldsymbol q}_0)=0.   \label{nh4} 
\end{eqnarray}
The auxiliary dynamical variable $h$ in these equations is fixed from the requirement of their consistency with the nonholonomic constraint (\ref{nh2}) as follows. The constraint (\ref{nh2}) must be satisfied at any time; this implies $dD/dt=0$. Using Eq. (\ref{nh3}) in this expression,  we get $h$ through ${\boldsymbol q}$ and $\dot {\boldsymbol q}$
\begin{eqnarray}\label{nh5}
h({\boldsymbol q}, \dot {\boldsymbol q})=\triangle^{-1}\left(\frac{\partial D}{\partial \dot q^i}\frac{\partial U}{\partial q^i}-
\frac{\partial D}{\partial q^i}\dot q^i\right), \qquad \mbox{where} \quad 
\triangle\equiv \frac{\partial D}{\partial \dot q^i} \frac{\partial D}{\partial \dot q^i}. 
\end{eqnarray}
For equations (\ref{nh3}) with this $h({\boldsymbol q}, \dot {\boldsymbol q})$, the function $D({\boldsymbol q}, \dot {\boldsymbol q})$ represents an integral of the motion. Therefore, all solutions with initial data (\ref{nh4}) will satisfy the nonholonomic constraint (\ref{nh2}).

The search for a variational problem and a Hamiltonian formulation of the Lagrange-d'Alembert equations (\ref{nh3}) is a problem with a century-long history, see the works \cite{Cha_2002, Cha_2008, Arn_2, Blo_2015} and references therein. Even in the case of linear and homogeneous nonholonomic constraints $A_i({\boldsymbol q})\dot q^i=0$, the problem is currently under active investigation, 
see \cite{Blo_2009, Sim_2025, Mes_2024} and references therein. A pseudo-Hamiltonian formulation (with the brackets without the Jacobi identity) for this case was suggested and discussed in the work \cite{Sch1994}. 

In this note, we present a simple procedure for writing both the variational problem and Hamiltonian formulation of the Lagrange-d'Alembert equations with nonholonomic constraints of a general form.

{\bf Comment.} The most natural way to take into account the constraint (\ref{nh2})  could be to add it to the Lagrangian action (\ref{nh1}) using a Lagrangian multiplier $g$
\begin{eqnarray}\label{nh6}
S=\int dt ~ \frac12\dot {\boldsymbol q}^2-U({\boldsymbol q})-g D({\boldsymbol q}, \dot {\boldsymbol q}).  
\end{eqnarray}
The equations that follow from this action are known as vakonomic mechanics \cite{Arn_2}.
Variation of the action with respect to $g$ implies the constraint $D=0$, which thus arises as part of the extremum conditions of this variational problem.  This perfectly works \cite{Rub1957, Arn_1, AAD_13, AAD_Book_2024} for the holonomic constraints $G({\boldsymbol q})=0$, but seems not to be suitable for nonholonomic ones: the equations implied by the action (\ref{nh6}) are not equivalent to the Lagrange-d'Alembert equations \cite{Arn_2}. 
The reason why the nonholonomic constraints (\ref{nh2}) should be taken into account using namely the Lagrange-d'Alembert equations is as follows. Nonholonomic constraints typically arise as a limiting case 
in theories with friction or damping, in the limit of infinite friction impeding motion in a prescribed direction \cite{Nei_1967}. In those cases where the resulting solutions can be compared with those to (\ref{nh3}) and (\ref{nh6}), it turns out that it is the solutions to the Lagrange-d'Alembert equations that give the correct answer, and therefore adequately describe this limiting case.

Chaplygin's sleigh is one of the test examples illustrating this situation, see the Appendix.

\section{Variational problem and Hamiltonian formulation for the Lagrange-D'alembert equations.}

{\bf The universal variational problem for a given system of ordinary differential equations.} We begin with a remarkable observation first made in the work \cite{Str_2023}.  Consider the following system of equations
\begin{eqnarray}\label{nh7}
S^i({\boldsymbol q}, \dot{\boldsymbol q}, \ddot{\boldsymbol q})=0, \qquad i=1, 2, \ldots, n. 
\end{eqnarray}
By squaring each of them and summing them up, we get a non-negative function that can be used to construct the following action functional: 
\begin{eqnarray}\label{nh8}
S=\int dt ~ \frac e2 \sum_i (S^i)^2.    
\end{eqnarray}
It involves the auxiliary variable $e(t)$. By varying the action with respect to $e$, we obtain $\sum_i (S^i)^2=0$, which is possible only if the equations (\ref{nh7}) are satisfied. Variation with respect to the remaining variables $q^i$ will yield the expressions whose terms are proportional either to the equations (\ref{nh7}) or to their time derivatives, and therefore will not provide any additional restrictions on the variables of the theory. In particular, there are no restrictions on the variable $e(t)$, which therefore has an ambiguous dynamics. Therefore, it represents a gauge degree of freedom, that is, it is an unobservable quantity in the sense of Dirac formalism for constrained systems \cite{Dir_1950, GT, deriglazov2010classical}. This indicates that there is some local symmetry in the theory, which will be described below.
In the result, we have obtained a variational problem for the equations (\ref{nh7}). 

Let us apply this construction to the Lagrange-d'Alembert equations (\ref{nh3}). We write them in the form 
\begin{eqnarray}\label{nh9}
\ddot{\boldsymbol q}-{\bf F}({\boldsymbol q}, \dot{\boldsymbol q})=0, \qquad \mbox{where} ~  
F^i\equiv-\frac{\partial U({\boldsymbol q})}{\partial q^i}+h({\boldsymbol q}, \dot{\boldsymbol q})
\frac{\partial D({\boldsymbol q}, \dot{\boldsymbol q})}{\partial \dot q^i},   
\end{eqnarray}
with $h({\boldsymbol q}, \dot{\boldsymbol q})$ given in Eq. (\ref{nh5}). Then the corresponding action reads
\begin{eqnarray}
S=\int dt ~ \frac e2\left(\ddot{\boldsymbol q}-{\bf F}({\boldsymbol q}, \dot{\boldsymbol q})\right)^2= \label{nh10} \qquad \\
\int dt ~ \frac e2\left[\dot{\bf v}-{\bf F}({\boldsymbol q, \bf v})\right]^2+{\boldsymbol\lambda}(\dot{\boldsymbol q}-{\bf v}), \label{nh11} 
\end{eqnarray}
where ${\boldsymbol\lambda}{\bf v}\equiv\sum_i\lambda_iv_i$, and so on. 
In the passage from (\ref{nh10}) to (\ref{nh11}), we follow the Ostrogradsky prescription \cite{Ost_1850, deriglazov2010classical}, introducing auxiliary variables ${\bf v}(t)$
and ${\boldsymbol\lambda}(t)$ to eliminate second derivatives in the action functional. The functional (\ref{nh11}) does not contain higher derivatives, so we can now construct a Hamiltonian formulation of our theory following the standard rules of the Dirac formalism. 

The variational problem (\ref{nh11}) is equally suitable for
cases of conservative forces $F^i({\boldsymbol q})=\partial U({\boldsymbol q})/\partial q^i$, and for dissipative (frictional) forces ${\bf F}({\boldsymbol q}, \dot{\boldsymbol q})$.

{\bf Comment.} There are two known possibilities to construct a variational problem for the harmonic oscillator with friction: $\ddot x-\omega^2 x+k^2\dot x=0$. First, this equation follows from the time-dependent Lagrangian $L=\frac 12e^{k^2t}(\dot x^2+\omega^2x^2)$. Second, a time-independent Lagrangian was suggested and discussed in the work \cite{Cha2007}. Our variational problem provides a third possibility and is suitable in the case of friction-like forces of general form ${\bf F}({\boldsymbol q}, \dot{\boldsymbol q})$.

{\bf Hamiltonian formulation of Lagrange-d'Alembert equations.} To obtain the Hamiltonian formulation of the theory (\ref{nh11}), we 
consider $6n+2$\,-dimensional phase space with the conjugate momenta $p_A=({\bf p}, {\boldsymbol\pi}, {\boldsymbol\pi}_\lambda, \pi_e)$ introduced for all configuration-space variables $z_A=({\bf q}, {\bf v}, {\boldsymbol\lambda}, e)$. They obey the canonical Poisson brackets $\{z_A, p_B\}=\delta_{AB}$. Temporal evolution of the conjugate momenta is determined by 
the standard rule $p_A=\partial L/\partial\dot z_A$. These expressions allow us to represent the velocities $\dot{\bf v}$ through ${\boldsymbol\pi}$
\begin{eqnarray}\label{nh12}
\pi^i=\frac{\partial L}{\partial\dot v^i}=e(\dot v^i-F^i), \qquad \mbox{then} \quad  \dot{\bf v}=
\frac 1e {\boldsymbol\pi}+{\bf F}. 
\end{eqnarray}
The remaining expressions for the momenta turn out to be primary constraints of the theory
\begin{eqnarray}\label{nh13}
\pi_e=0, \qquad {\boldsymbol\pi}_\lambda=0, \qquad {\bf p}-{\boldsymbol\lambda}=0. 
\end{eqnarray}
The explicit form of the complete Hamiltonian $H_c=p_A\dot z_A-L+\mu_e\pi_e+{\boldsymbol\mu}_\lambda{\boldsymbol\pi}_\lambda+
{\boldsymbol\mu}_q({\bf p}-{\boldsymbol\lambda})$ is as follows:
\begin{eqnarray}\label{nh14}
H_c=\frac{1}{2e}{\boldsymbol\pi}^2+{\boldsymbol\pi}{\bf F}({\boldsymbol q}, {\bf v})+ {\boldsymbol\lambda}{\bf v}+
\mu_e\pi_e+{\boldsymbol\mu}_\lambda{\boldsymbol\pi}_\lambda+{\boldsymbol\mu}_q({\bf p}-{\boldsymbol\lambda}). 
\end{eqnarray}
The Lagrangian multipliers for the primary constraints were denoted by $\mu_e$, ${\boldsymbol\mu}_\lambda$, and ${\boldsymbol\mu}_q$.  Hamiltonian equations of the theory read as follows:  $\dot z_A=\{ z_A, H_c\}$, $\dot p_A=\{ p_A, H_c\}$.

We should now reveal all higher-stage constraints and determine the Lagrangian multipliers from the condition of conservation of the primary constraints over time. Following Dirac's procedure \cite{Dir_1950, GT, deriglazov2010classical}, we obtain
\begin{eqnarray}
\pi_e=0, \quad \Longrightarrow \quad {\boldsymbol\pi}=0, \quad \Longrightarrow \quad {\boldsymbol\lambda}=0, \quad \Longrightarrow \quad 
{\boldsymbol\mu}_\lambda=0; \label{nh15} \\
{\boldsymbol\pi}_\lambda=0, \quad \Longrightarrow \quad {\boldsymbol\mu}_q={\bf v}; \label{nh16} \\
{\bf p}-{\boldsymbol\lambda}=0,  \quad \Longrightarrow \quad 
{\boldsymbol\pi}\frac{\partial {\bf F}({\boldsymbol q}, {\bf v})}{\partial q^i}-\mu_{\lambda}^i=0. \label{nh17} 
\end{eqnarray}
These equalities reveal the final form of all constraints: 
\begin{eqnarray}
\pi_e=0, \qquad {\boldsymbol p}=0, \qquad {\boldsymbol\pi}=0; \label{nh18} \\ 
{\boldsymbol\lambda}=0, \qquad {\boldsymbol\pi}_\lambda=0, \label{nh19}
\end{eqnarray}
as well as determine most of the Lagrangian  
multipliers: ${\boldsymbol\mu}_\lambda=0$, and ${\boldsymbol\mu}_q={\bf v}$. The multipliers obtained can be substituted into the Hamiltonian, which then reads as follows: 
\begin{eqnarray}\label{nh20}
H=\frac{1}{2e}{\boldsymbol\pi}^2+{\boldsymbol\pi}{\bf F}({\boldsymbol q}, {\bf v})+{\bf p}{\bf v}+\mu_e\pi_e.
\end{eqnarray}
Let us discuss the equations of motion of the theory. 
The pair of conjugate auxiliary variables ${\boldsymbol\lambda}$  and ${\boldsymbol\pi}_\lambda$ has trivial 
dynamics: ${\boldsymbol\lambda}=0$, ${\boldsymbol\pi}_\lambda=0$, and can now be omitted from the formalism\footnote{Formally, we should construct the Dirac bracket of these constraints, after which they can be used as strong equalities.}. The explicit form of Hamiltonian 
equations $\dot{\boldsymbol q}=\{ {\boldsymbol q}, H\}$,  $\dot{\boldsymbol v}=\{ {\boldsymbol v}, H\}$ and $\dot e=\{e, H\}$ is as follows:  
\begin{eqnarray}
\dot{\boldsymbol q}={\bf v}, \qquad \dot{\boldsymbol v}={\bf F}({\boldsymbol q}, {\bf v});  \label{nh21} \\
\dot e=\mu_e. \label{nh22}
\end{eqnarray}
Hamiltonian equations for the remaining variables $\pi_e, {\boldsymbol p}, {\boldsymbol\pi}$ turn out to be satisfied as a consequence of the constraints (\ref{nh18}). 

The variables ${\boldsymbol q}$ and ${\bf v}$ have unambiguous dynamics. Their first-order equations (\ref{nh21}) are equivalent to the original Lagrange-d'Alembert equations (\ref{nh9}). The multiplier $\mu_e$ was not determined in the course of the Dirac procedure, so the equation (\ref{nh22}) implies that the auxiliary variable $e(t)$ has an arbitrary dynamics, and hence it represents an unobservable (gauge) degree of freedom of the Dirac formalism. The equation itself is invariant with respect to the following infinitesimal local symmetry with the parameter $\alpha(t)$: $\delta e=\alpha$, $\delta\mu_e=d\delta e/dt$.

Our final Hamiltonian (\ref{nh20}) can be used to construct the first-order (Hamiltonian) variational problem of our theory  
\begin{eqnarray}\label{nh23}
S_H=\int dt ~ {\bf p}\dot{\boldsymbol q}+{\boldsymbol\pi}\dot{\bf v}+\pi_e\dot e-\frac{1}{2e}{\boldsymbol\pi}^2-{\boldsymbol\pi}{\bf F}({\boldsymbol q}, {\bf v})-{\bf v}{\bf p}-\mu_e\pi_e.
\end{eqnarray}
Variation of the action with respect to all variables involved implies the equations (\ref{nh18}), (\ref{nh21}) and (\ref{nh22}), that is the complete system of equations of our theory. 

\section{Conclusion.}

In the present work, we have shown that there are no ``non-Lagrangian" equations in classical mechanics: any given system of ordinary differential equations turns out to be the extremum condition of the universal variational problem (\ref{nh8}). For the equations of the form $\ddot{\boldsymbol q}-{\bf F}({\boldsymbol q}, \dot{\boldsymbol q})=0$, where the force ${\bf F}({\boldsymbol q}, \dot{\boldsymbol q})$ is a given function of positions and velocities, we have constructed the corresponding Hamiltonian formulation. When ${\bf F}$ is of special form (\ref{nh9}), this gives the Hamiltonian formulation of Lagrange-d'Alembert equations with nonholonomic constraint $D({\boldsymbol q}, \dot{\boldsymbol q})=0$. The generalization to the case of several nonholonomic constraints $D_\alpha({\boldsymbol q}, \dot{\boldsymbol q})=0$ is straightforward. 

In its final form, the obtained Hamiltonian formulation can be described as follows. Presenting the original equations $\ddot{\boldsymbol q}-{\bf F}({\boldsymbol q}, \dot{\boldsymbol q})=0$ in the first-order form $\dot{\boldsymbol q}={\bf v}$, 
$\dot{\boldsymbol v}={\bf F}({\boldsymbol q}, {\bf v})$, their Hamiltonian formulation consist of the unique primary constraint $\pi_e=0$, the Hamiltonian 
\begin{eqnarray}\label{nh25}
H=\frac{1}{2e}{\boldsymbol\pi}^2+{\boldsymbol\pi}{\bf F}({\boldsymbol q}, {\bf v})+{\bf p}{\bf v}+\mu_e\pi_e,
\end{eqnarray}
and the canonical Poisson brackets $\{q^i, p_j\}=\delta^{i}{}_j$, $\{v^i, \pi_j\}=\delta^{i}{}_j$, and $\{e, \pi_e\}=1$ defined on $4n+2$\,-dimensional phase space. This gives the equations (\ref{nh18}), (\ref{nh21}) and (\ref{nh22}). As was shown in the previous section, they represent the Hamiltonian form of the Lagrange-d'Alembert equations. 

Equivalently, these equations turn out to be the extremum conditions of the first-order action functional (\ref{nh23}). 

Let us conclude with two comments about some special properties of the obtained formulation. \par

\noindent {\bf 1.} Similarly to the case of a reparametrization-invariant theory, the Hamiltonian (\ref{nh25}) of our theory  vanishes on the surface of constraints (\ref{nh18}).  \par

\noindent {\bf 2.} The Lagrangian action (\ref{nh10}) led us to the Hamiltonian (\ref{nh25}) which contains the terms $(\pi^i)^2$, $i=1, 2, \ldots, n$, quadratic in the Dirac constraints $\pi^i=0$.  It should be noted that some general statements \cite{GT} on the structure of a constrained theory do not apply to the case of Hamiltonians with quadratic constraints. In particular, note that all Poisson brackets between the constraints (\ref{nh18}) vanish, meaning that all $2n+1$ are first-class constraints of Dirac formalism. In the case of a standard Hamiltonian, all conjugate variables to the constraints would have ambiguous dynamics, leading to a theory in which there is no physical sector at all.
However, the variables $q^i$ and $v^i$ of our theory have unambiguous dynamics according to the equations (\ref{nh21}), and thus represent the observable quantities. A more detailed discussion of these issues can be found in \cite{AAD_2021.1}, where a Hamiltonian with quadratic constraints has been used to construct a massless polarized particle with null geodesics strictly protected against the hypothetical Magnus and spin Hall effects of light.

\begin{acknowledgments}
I am grateful to Dr. W. Struyve for useful discussions that allowed me to correct inaccuracies in the work.
\end{acknowledgments}

\section{Appendix.}

In this appendix, we compare solutions to the equations of Chaplygin sleigh in four different cases and verify that the Lagrange-d'Alembert equations, unlike the vakonomic equations, correctly describe the Chaplygin sleigh in the limit of infinite friction.

Consider Chaplygin's symmetrical sleigh \cite{Cha_2008}, which is a rigid body of inertia $I$ and mass $m$ that rests in a plane at three points. Two of them are simple, freely sliding legs, and the third is the point of contact of a small, sharp-rimmed wheel whose horizontal axis is fixedly attached to the body. The center of mass lies on the wheel axis. Dynamical variables of the problem are the center-of-mass coordinates ${\bf y}(t)=(y^1(t), y^2(t))$, and the angle $\varphi(t)$ between the wheel's plane and the horizontal axis of the Laboratory system. If we assume that the wheel also slides freely on the plane, the Lagrangian of the theory will be as follows: $L_0=\frac m2 \dot{\bf y}^2+ \frac I2 \dot\varphi^2$.  Without loss of generality, we take the initial data as follows:
\begin{eqnarray}\label{nh26}
y^1(0)=y^2(0)=0, \qquad \dot y^1(0)=v^1_0,  \quad \dot y^2(0)=0, \qquad \varphi(0)=0, \quad \dot\varphi(0)=\omega. 
\end{eqnarray}
{\bf 1. Chaplygin's symmetrical sleigh with friction.}  Assume that in the case of the wheel sliding in the direction normal to its plane, the sleigh experiences a friction force with the friction coefficient $k$. Then equations of motion read as follows
\begin{eqnarray}\label{nh27}
\ddot\varphi=0, \quad \mbox{then} \quad \varphi=\omega t; \qquad \ddot y^1=\frac kmv^2\sin\varphi ;  \qquad 
\ddot y^2=-\frac{k}{m}v^2\cos\varphi. 
\end{eqnarray}
Here and below we denote by $v^i$ components of velocity relative to the body-fixed frame  
\begin{eqnarray}\label{nh28}
v^1\equiv \dot y^1\cos\varphi+\dot y^2\sin\varphi, \qquad v^2\equiv -\dot y^1\sin\varphi+\dot y^2\cos\varphi. 
\end{eqnarray}
In the case of strong friction $k>2\omega m$, the solution of the problem (\ref{nh27}), (\ref{nh26}) is $\varphi=\omega t$ and 
\begin{eqnarray}\label{nh29}
y^1(t)=y^1_{\infty}+\frac{v^1_0}{\omega(d_2-d_1)}\left\{d_2e^{-d_1t}\sin(\omega t+\phi_1)-d_1e^{-d_2t}\sin(\omega t+\phi_2)\right\}, \cr
y^2(t)=y^2_{\infty}-\frac{v^1_0}{\omega(d_2-d_1)}\left\{d_2e^{-d_1t}\cos(\omega t+\phi_1)-d_1e^{-d_2t}\cos(\omega t+\phi_2)\right\},
\end{eqnarray}
where $d_{2, 1}\equiv (k\pm\sqrt{k^2-4\omega^2 m^2})/2m$. 
The integration constants $y^1_{\infty}=2 v^1_0m/(\omega k)$ and $y^2_{\infty}=v^1_0/\omega$ determine the final position of the  sleigh as $t\rightarrow\infty$, while the phases turn out to be $\sin\phi_1=-2d_1/(\omega^2+d^2_1)$ and  $\sin\phi_2=-2d_2/(\omega^2+d^2_2)$. 

In the limit of infinite friction, $k\rightarrow\infty$, the solution  (\ref{nh29}) turns into 
\begin{eqnarray}\label{nh30}
\varphi(t)=\omega t, \qquad y^1(t)=\frac{v^1_0}{\omega}\sin\omega t, \qquad y^2(t)=\frac{v^1_0}{\omega}-\frac{v^1_0}{\omega}\cos\omega t. 
\end{eqnarray}
The center of mass moves with a constant angular velocity along a circle of radius $\frac{v^1_0}{\omega}$, while the axis of the wheel always is directed toward the center of this circle.

{\bf 2. Lagrange-d'Alembert equations of Chaplygin's symmetrical sleigh with linear nonholonomic constraint.} Now let us assume that the wheel cannot slide in the direction normal to its plane, and take this into account with the help of linear nonholonomic constraint
\begin{eqnarray}\label{nh31}
\dot y^1\sin\varphi-\dot y^2\cos\varphi=0. 
\end{eqnarray}
The Lagrange-d'Alembert equations (\ref{nh3}) for this case read as follows
\begin{eqnarray}\label{nh32}
I\ddot\varphi=0, \quad \mbox{then} \quad \varphi=\omega t; \qquad \ddot y^1=-v^1\omega\sin\varphi;  \qquad 
\ddot y^2=v^1\omega\cos\varphi. 
\end{eqnarray}
Although they are different from (\ref{nh27}), their solution with the initial data (\ref{nh26}) is given by formulas (\ref{nh30}). 

{\bf 3. Lagrange-d'Alembert equations of Chaplygin's symmetrical sleigh with nonlinear nonholonomic constraint.} Here we test a nonlinear version of the constraint (\ref{nh31}) of the following form: 
\begin{eqnarray}\label{nh33}
\frac{\dot y^2}{\dot y^1}-\tan\varphi=0. 
\end{eqnarray}
The Lagrange-d'Alembert equations (\ref{nh3}) for this case read as follows
\begin{eqnarray}\label{nh34}
I\ddot\varphi=0, \quad \mbox{then} \quad \varphi=\omega t; \qquad \ddot y^1=-\omega \dot y^2;  \qquad 
\ddot y^2=\omega\dot y^1. 
\end{eqnarray}
Although they are different from (\ref{nh27}) and (\ref{nh32}), their solution with the initial data (\ref{nh26}) is given by formulas (\ref{nh30}). 

{\bf 4. Vakonomic mechanics for Chaplygin's symmetrical sleigh.} For the case, the action (\ref{nh6}) reads
\begin{eqnarray}\label{nh35}
S=\int ~ \frac m2 \dot{\bf y}^2+ \frac I2 \dot\varphi^2-g(-\dot y^1\sin\varphi+\dot y^2\cos\varphi). 
\end{eqnarray}
This can be analysed in the Dirac formalism for constrained systems, and leads to the following equation for determining the angle $\varphi(t)$
\begin{eqnarray}\label{nh36}
2Im\ddot\varphi=[c^2-(mv_0^1)^2]\sin2\varphi+cmv_0^1 \cos2\varphi,
\end{eqnarray}
where $c$ is an integration constant that appears in the process of solving equations for conjugate momenta. For any choice of this constant, this equation does not reduce to the correct equation $\ddot\varphi=0$. Thus, vakonomic mechanics does not describe Chaplygin's sleigh in the limit of infinite friction.

\end{document}